\documentclass[twocolumn,letterpaper]{IEEEAerospaceCLS}  %

\usepackage[]{graphicx}
\usepackage{booktabs}
\usepackage{tabularx}
\usepackage{amsmath}
\usepackage{xcolor}
\usepackage[colorlinks=true, urlcolor=blue, linkcolor=black, citecolor=black]{hyperref}

\usepackage{booktabs}
\usepackage{multirow}
\usepackage{threeparttable}
\usepackage{pifont}
\usepackage{array}
\usepackage{graphicx}
\newcommand{\cmark}{\ding{51}}                 %
\newcommand{\pmark}{(\ding{51})}               %
\newcommand{\xmark}{--}                        %
\newcommand{\na}{\textit{n/a}}                 %
\newcommand{\ignore}[1]{}
\graphicspath{{./}}

\usepackage{colortbl}
\definecolor{tblgroup}{gray}{0.92}

\begin{document}

\title{Benchmarking Open-Source O-RAN and 5G NTN Stacks on Embedded ARM Platforms for Space Payloads}

\author{%
\mbox{Esteban Municio$^{1}$},
\mbox{Oscar Baselga$^{1,2}$},
\mbox{Oscar Lasierra$^{1}$},
\mbox{Gines Garcia-Aviles$^{1}$},
\mbox{Xavier Costa-P\'erez$^{1,3,4}$}\\[1.1ex]
$^{1}$i2CAT Foundation, Barcelona, Spain\\[0.5ex]
$^{2}$Universitat Politècnica de Catalunya - BarcelonaTech (UPC), Barcelona, Spain \\
$^{3}$NEC Laboratories Europe, Heidelberg, Germany\qquad\\
$^{4}$ICREA, Barcelona, Spain \\
 \{esteban.municio,oscar.baselga,oscar.lasierra,gines.garcia,xavier.costa\}@i2cat.net
}

\maketitle

\raggedbottom
\widowpenalty=10000 \clubpenalty=10000

\thispagestyle{plain}
\pagestyle{plain}

\begin{abstract}
Regenerative satellite payloads, which host the 5G Next Generation Node B (gNB) onboard instead of transparently relaying the radio signal, are a key building block of the Non-Terrestrial Network (NTN) integration that 3GPP started in Release 17. Combining them with the Open Radio Access Network (O-RAN) architecture adds software-defined control through RAN Intelligent Controllers (RICs), enabling in-orbit reconfiguration, multi-tenancy, and data-driven radio resource management. Open-source implementations make this vision accessible to research institutions and New Space missions, yet there is little quantitative evidence on how they behave on the embedded processors that a satellite can actually carry. This paper presents an out-of-the-box experimental benchmark of two open-source gNB stacks (i.e., OpenAirInterface/Duranta and srsRAN/OCUDU), and two Near-Real-Time RICs (i.e., the ORAN-SC Near-RT RIC and FlexRIC). All components are deployed on an ARMv8 platform representative of the processing units used in small-satellite payloads, driving a USRP B210 software-defined radio. We compare them in terms of functional compliance (3GPP NTN features, O-RAN interfaces, multi-tenancy), resource efficiency, real-time behavior, and engineering suitability. Results show that these platforms can host a joint gNB + near-RT RIC deployment using any of the considered components, although further optimizations are still required for carrier-grade performance. However, the performance and standards compliance vary significantly among them. Whereas srsRAN has a smaller memory footprint, OAI shows a lower CPU utilization. And while both have significant NTN feature support, they still differ in some relevant functionalities. 
On the controller side, FlexRIC requires far fewer resources than the ORAN-SC RIC, while the latter offers broader O-RAN compliance. Building on these results, we derive a decision framework that maps mission profiles and Size, Weight, and Power (SWaP) budgets to gNB/near-RT RIC combinations. Finally, we discuss the open gaps remaining before these stacks can be flown in NTN regenerative payloads.

\end{abstract}

\tableofcontents

\section{Introduction}

Satellite systems are becoming an integral part of mobile networks. %
Since Release 17, 3GPP has specified support for 5G New Radio (NR) over Non-Terrestrial Networks (NTN), addressing the long propagation delays, large Doppler shifts, and dynamic cell management characteristic of satellite gNBs~\cite{3gpp_tr38821,3gpp_ts38300,lin2021fivegspace}. The first NTN deployments standardized by 3GPP in Release 17 rely on transparent payloads, where the satellite acts as a bent-pipe relay, and the gNB remains on the ground. The regenerative payload concept was then standardized in Release 19, proposing hosting the gNB or part of it onboard. This shortens the radio control loops and decouples them from the feeder-link availability, paving the way for inter-satellite routing and in-orbit edge processing~\cite{giordani2021ntn}.

Hosting the gNBs onboard also raises the question of how the Radio Access Network (RAN) is operated in space. The Open RAN (O-RAN) architecture already answers this question for terrestrial networks by disaggregating the gNB through open interfaces and exposing it to RICs~\cite{polese2023oran}. Applied to a regenerative satellite, %
O-RAN enables three major levers: flexible reconfiguration of network functions after launch; sharing of the gNB payload among third parties, either as tenants (multi-tenancy) or as providers of xApps; and fast closed-loop radio resource management and mobility optimization that adapts to orbital dynamics.%
Where the Radio Intelligent Controllers (RICs) run is equally important. The Non-Real-Time RIC (Non-RT RIC) sets policies through the A1 interface on time scales above one second, which tolerates the space--ground link, whereas the Near-Real-Time RIC (Near-RT RIC) controls the RAN through the E2 interface with loops between 10\,ms and 1\,s. For a geostationary satellite, the payload--gateway round trip alone  may take about a quarter of a second~\cite{3gpp_tr38821}, and a low Earth orbit satellite sees a given gateway for only a few minutes per pass. 
Closing near-real-time loops from the ground is therefore impractical, which motivates hosting the Near-RT RIC onboard next to the gNB, while the Non-RT RIC remains on the ground.

Open-source projects have become the de facto platforms to prototype such systems. OpenAirInterface (OAI)~\cite{oai} (now Duranta) and srsRAN Project~\cite{srsran} (now OCUDU) provide complete 5G gNB implementations with NTN extensions and E2 agents, while FlexRIC~\cite{flexric} and the O-RAN Software Community (ORAN-SC) Near-RT RIC~\cite{oransc} provide two very different Near-RT RIC implementations: a lightweight, monolithic software development kit and a cloud-native reference platform composed of microservices, respectively. These projects are, however, usually developed and evaluated on x86 servers with abundant computing, memory, and power resources~\cite{rouili2024evaluating,bonati2023openrangym}. Satellite payload processors are embedded Systems-on-Chip, typically based on ARM cores, with a few cores, limited memory and storage, and a strict power budget that must be shared with the rest of the payload software~\cite{george2018onboard}. 
While some recent works have tested OAI to study different NTN architectures~\cite{baselga2026empirical,volk2022field}
, to the best of our knowledge, no work has yet compared alternative gNB stacks and RIC platforms, and their combinations, on the same embedded hardware, nor quantified the cost of hosting the Near-RT RICs onboard the satellite.

This paper addresses this gap by out-of-the-box benchmarking OAI/Duranta and srsRAN Project/OCUDU as gNBs, and the ORAN-SC Near-RT RIC and FlexRIC as Near-RT RICs, on an AMD Xilinx Zynq UltraScale+ ZCU104 board, whose MPSoC family is widely used in small-satellite payload computers~\cite{munoz2025towards}. The board drives a USRP B210 software-defined radio and runs the software in containers on a real-time Linux kernel. The evaluation is guided by four questions: (i) whether each component fits within the resources of an embedded payload processor, (ii) which performance can be sustained without applying major modifications and what limits it, (iii) which 3GPP NTN and O-RAN support each component offers, and (iv) which gNB--RIC combination best suits a given mission profile. In this sense, the main contributions of this paper are:
\begin{enumerate}
    \item[•] {An out-of-the-box experimental benchmark of two open-source gNB stacks and %
    Near-RT RICs compliant with 3GPP and O-RAN standards on embedded ARMv8 hardware}, including the qualitative analysis of each of the design choices.
    \item[•] {A joint gNB--RIC study}, assessing the aggregated footprint and the interoperability of each combination when deployed jointly on the platform.
    \item[•] {A decision framework} that maps mission profiles and SWaP budgets to recommended gNB-RIC combinations, together with the open gaps still remaining.
\end{enumerate}

This paper is organized as follows. Section~\ref{sec:background} introduces O-RAN-enabled NTN and the evaluated gNB and RIC candidates. Section~\ref{sec:methodology} describes the experimental setup. Sections~\ref{sec:results} and~\ref{sec:ric_results} present the experimental evaluation of the gNBs and the Near-RT RICs. Section~\ref{sec:decision_guide} discusses the results and derives the selection framework, and Section~\ref{sec:conclusion} concludes the paper.

\section{Background}
\label{sec:background}

\subsection{O-RAN Enabled NTN}
NTNs are communication networks that use either airborne platforms or satellites that carry relay nodes or base stations. 3GPP Release~17 introduced the first normative support of 5G NR over NTN and has been extended in subsequent Release 18 and Release 19. Satellite links challenge the original NR design by adding significant one-way propagation delays (from a few milliseconds in LEO to more than $250\,\text{ms}$ in GEO), 
introducing Doppler shifts of up to ${\pm}48\,\text{kHz}$ %
for LEO with a $2\,\text{GHz}$ (S-band) carrier~\cite{3gpp_tr38821}, 
and adding cell coverage dynamics that move as the satellite passes by. The specification addresses these effects through: a) broadcasting satellite ephemeris and common timing information in SIB19, which helps the UE to locate the satellite~\cite{3gpp38331}, b) pre-compensating delay and Doppler at the UE using their relative position and the ephemeris~\cite{3gpp_ts38300}, extending the scheduling timing relationships with a
cell-specific offset $K_\mathrm{offset}$, plus $K_\mathrm{mac}$ when the downlink and uplink frame timing are not aligned at the gNB~\cite{3gpp38213}; and
c) offsetting MAC timers by the UE--gNB round-trip time and allowing HARQ feedback to be disabled per process~\cite{3gpp38321}. 
Still, NTN standardization is ongoing. The 6G studies started in Release~20 consider NTN an integral part of the system, with multi-layer architectures integrating terrestrial, aerial and satellite segments~\cite{rago2025innovative}.

NTN defines two payload architectures, a transparent payload, which only performs frequency conversion and amplification, keeping the complete gNB on the ground, and a regenerative payload that processes the 5G NR signals onboard, hosting either the complete gNB or only the DU, with the CU on the ground connected through F1 over the feeder link~\cite{3gpp_tr38821,baselga2026empirical}. Moving from transparent payload approaches to regenerative payload approaches aims to move network functionalities to space, such as enabling Inter-Satellite Links (ISLs) or allowing base stations to operate regardless of whether or not they have direct visibility with a ground station.%

O-RAN provides the architectural blueprint to realize this. 
Driven by the O-RAN Alliance, it extends the 3GPP RAN with open interfaces and programmable control~\cite{polese2023oran,oran_oad}. O-RAN considers a gNB disaggregated into a Central Unit (O-CU), a Distributed Unit (O-DU), and a Radio Unit (O-RU) interconnected through the F1-c/F1-u and Open Fronthaul interfaces. Two controllers then close loops at different time scales: i) the Non-Real-Time RIC, hosted in the Service Management and Orchestration framework and issues policies through the A1 interface, and ii) the Near-RT RIC, which runs xApps %
for intelligent monitoring and control of the RAN through the E2 interface and its corresponding service models E2SMs, such as the E2SM-KPM for performance monitoring and the E2SM-RC for radio control. Finally, O-RAN also includes the O1 and O2 interfaces for the management of the infrastructure.

This model is attractive for a regenerative payload. A satellite cannot be easily serviced after launch, so updating xApps (e.g., a new ML-based approach for optimizing beamforming or an algorithm that optimizes resources based on real-time demands) extends the useful life of the payload. Also, disaggregation allows for getting the most of the SWaP satellite platform by relocating network functions depending on the link budgets. In addition to this, orbital motion is dynamic yet known in advance from the ephemeris, which makes scheduling, handover, or inter-satellite load balancing natural targets for the closed-loop control of a near-RT RIC. Also, open interfaces supporting, for example, multi-tenancy via a Multi-Operator Core Network (MOCN) model help amortize the cost of the space segment across several operators. These benefits were nevertheless conceived for terrestrial deployments on commodity servers with abundant cores, memory, and power. A payload processor is instead an embedded System-on-Chip with a handful of ARM cores, a few gigabytes of memory and limited wattage.
Whether current open-source implementations can operate under these conditions, and at what cost, is the question we address next.

\subsection{5G gNB Candidate Stacks}

The open-source 5G RAN ecosystem has consolidated around two primary technology lineages, both of which have recently transitioned to Linux Foundation governance \cite{linuxfundation,ocudu_mig}. The first lineage originates from Software Radio Systems (SRS) under the srsRAN Project, while the second originates from EURECOM as OAI. They present different alternatives regarding the nature of the contributions to the codebase, the implementation roadmap of the NTN feature set, and the licensing regime.

\subsubsection{srsRAN Project / OCUDU}
Developed originally by SRS, the srsRAN Project provides a modular, high-performance open-source 5G RAN implementation engineered in C++ for modern multi-core architectures~\cite{srsran}. It offers an optimized, clean, and thread-safe implementation of 5G NR control-plane and user-plane protocols, encompassing SDAP, RRC, PDCP, RLC, MAC, and PHY. The framework features a complete gNB implementation that supports functional splits into CU and DU blocks, specifically Split 2, Split 7.2, and Split 8. To ensure low latency and execution efficiency, srsRAN leverages modern C++ abstractions alongside SIMD vectorization, including AVX2 for x86 and NEON for ARM processors. For NTN operations, srsRAN supports foundational 3GPP Release 17 features tailored mostly for GEO scenarios. Its capabilities include broadcasting the SIB19 (ECEF or orbital ephemeris, epochTime, ta-Common and drifts) and the cell-specific K\_offset and T430. Also, its scheduler supports 32 HARQ processes and UL HARQ mode B among others, but enabling them for a UE, requires a closed enterprise build.
Furthermore, the framework integrates an embedded E2 Agent that supports key O-RAN service models, such as E2SM-KPM and E2SM-RC. The open-source continuation of this lineage under Linux Foundation governance is known as the OCUDU project, which has the permissive BSD-3 Clause licence and serves as the reference implementation for O-RAN-compliant O-CU and O-DU protocol stacks, directly inheriting srsRAN's architectural design, C++ codebase, and runtime execution model. Regarding NTN, OCUDU adds a larger support including DL HARQ feedback disabling, TA reporting, conditional handover or satellite switch with re-synchronization.

\subsubsection{OpenAirInterface (OAI) / Duranta}
Maintained by the EURECOM-led OAI Software Alliance, OAI is one of the most established open-source wireless software platforms within the research community. It is widely utilized as an experimental platform for 3GPP feature prototyping due to its broad community support. OAI distinguishes itself by providing a full-stack 5G NR implementation that covers the gNB, the 5GC, and the UE~\cite{oai}. Architecturally, OAI supports the standardized gNB deployments, including functional split (CU and DU) and monolithic gNB configurations, and implements an embedded E2 Agent capable of exposing fine-grained telemetry and control across the MAC, RLC, PDCP, and RRC layers. This E2 Agent supports E2SM-RC, E2SM-KPM, and customized E2 service models. 

OAI implements the Rel-17 NTN essentials and emerging Rel-18 NTN features for GEO, MEO, and LEO satellites both at the gNB and the UE. This includes using a SIB19 with the ECEF state-vector ephemeris, the \texttt{ta-Common} drift, the cell-specific $K_\mathrm{offset}$, and handover with \texttt{ntn-Config}. The OAI nrUE derives the timing advance from SIB19 and its own position and pre-compensates the UL frequency. This makes OAI the only one of the two stacks that can close an NTN link end to end without a commercial NR-NTN UE. To bridge the gap between academic research prototyping and commercial-grade industrial deployments, the Duranta initiative has been established within the OAI ecosystem, specifically targeting commercial-grade hardening and containerized deployment.

\subsection{Near-RT RIC Candidate Platforms}

The near-RT RIC operates at the edge of the RAN %
and enables the software-driven control of gNB through xApps.
Two primary open-source candidate platforms are available: the ORAN-SC near-RT RIC and FlexRIC.

\subsubsection{O-RAN SC Near-RT RIC}
It is the reference implementation maintained jointly by the O-RAN Alliance and the Linux Foundation~\cite{oransc}. Built as an enterprise-grade, cloud-native platform, the ORAN-SC RIC consists of a comprehensive suite of microservices running within Docker and Kubernetes containers (including the E2 Manager, E2 Termination, A1 Mediator, Subscription Manager, xApp Onboarding, and a Shared Data Layer utilizing a Redis DB). It provides architectural and native compliance with O-RAN specifications, supporting E2AP v3.0+, A1 policy interfaces and enforcement, as well as O1 configuration and management protocols. The platform offers robust isolation, high security, and multi-tenancy support enforced through Kubernetes namespaces and Role-Based Access Control (RBAC). However, its modular microservices architecture incurs noticeable container management and multi-pod execution overhead, alongside a wide attack surface. Consequently, this platform requires careful evaluation before deployment on constrained embedded hardware, such as onboard satellite processing units.

\subsubsection{FlexRIC}
Developed by EURECOM, FlexRIC is an ultra-lightweight, flexible, and open-source Near-RT RIC platform specifically designed for resource-constrained edge and embedded execution environments~\cite{flexric}. Built prioritizing modularity and efficiency, FlexRIC utilizes a specialized C-based SDK to implement E2 Termination points, xApps, and service models. The platform allows xApps (which can be written in either C or Python) to execute either as dedicated standalone processes or to be linked directly within the RIC memory space to minimize latency and overhead.

\begin{figure}
  \centering
  \includegraphics[width=\linewidth]{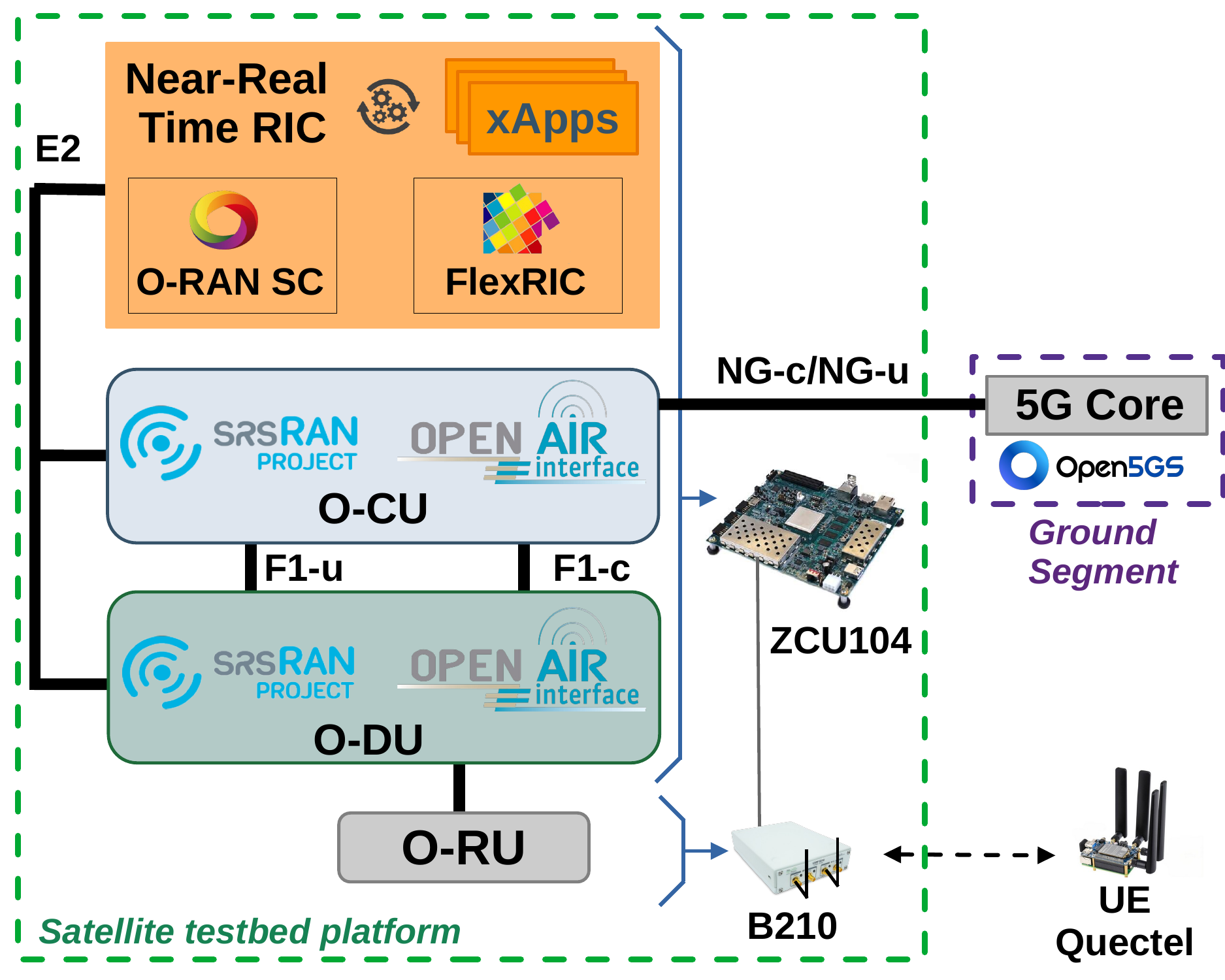}
  \caption{O-RAN Testbed used for benchmarking.}
  \label{fig:testbed}
\end{figure}

\begin{table}
  \renewcommand{\arraystretch}{1.2}
  \caption{\textbf{Software versions used in the experiments}}
  \label{tab:sw_versions}
  \centering
  \small
  \begin{tabular}{c | c}
  \toprule
  \textbf{Component} & \textbf{Version / commit} \\
  \midrule
  \hline
  OAI gNB               & v2.4.0 / 91e7030c \\
  srsRAN Project gNB   & main / 4bf1543 \\
  FlexRIC              & dev / ef6d722f \\
  ORAN-SC Near-RT RIC & i-release / 621ade262 \\
  \bottomrule
  \end{tabular}
\end{table}

FlexRIC supports standard O-RAN service models (such as E2SM-KPM and E2SM-RC) alongside customized, fine-grained, multi-layer service models capable of exposing MAC, RLC, and PDCP telemetry and control. Due to its minimal footprint, FlexRIC is well suited for onboard satellite processing where compute and memory resources are constrained. However, a notable trade-off is that it lacks out-of-the-box native support for standardized O1 (management and configuration) and A1 (policy) interfaces.

\begin{figure*}[htbp]
  \centering
  \includegraphics[width=\linewidth]{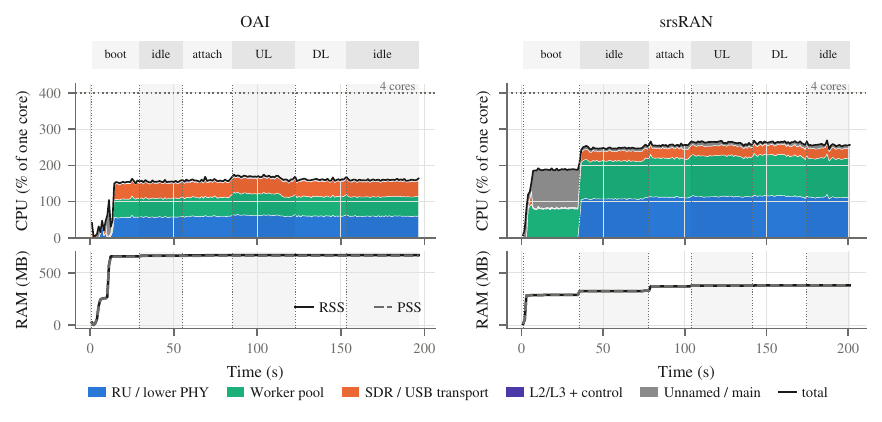}
  \caption{\textbf{gNB CPU and memory utilization comparison on the ZCU104 for both OAI and srsRAN.}}
  \label{fig:gnb_cpu}
\end{figure*}

\section{Methodology}
\label{sec:methodology}

\subsection{Hardware Platform}
\label{subsec:hw-platform}

The target board is an AMD Zynq UltraScale+ ZCU104 evaluation kit. Its MPSoC integrates a quad-core ARM Cortex-A53 (AArch64), programmable logic (FPGA fabric) and 2 GB of DDR4.  AMD's radiation-tolerant XQR line currently covers other families (e.g., Kintex UltraScale and Versal), but Zynq UltraScale+ MPSoCs already fly in commercial small-satellite on-board computers, including the 6GStarLab 6U CubeSat launched in November 2025 to support in-orbit NTN experimentation~\cite{munoz2025towards}. The ZCU104 is therefore a representative for such payload processors,  with four in-order cores sharing a 1MB L2 cache, with NEON as the only vector extension.

The ZCU104 drives an Ettus USRP B210 SDR over USB 3.0, operating over-the-air in a controlled area. The software runs on Linux 5.15 patched with PREEMPT\_RT, and every component is deployed as Open Container Initiative (OCI) container managed through Podman $5.0.3$, %
a feasible management option for the onboard resources. The 5G Core, based on Open5GS~\cite{open5gs}, and the commercial UE, a Quectel RM520N, are hosted off-board, so the measured footprint corresponds to the payload software alone. To keep runs comparable, all four cores are pinned at 1.2 GHz under the userspace governor, and idle states are disabled (\texttt{cpuidle.off=1}). Neither frequency scaling nor C-state residency therefore contributes to the differences reported below. The on-chip temperature reads ${\approx}36\,^{\circ}\text{C}$ throughout, far below any throttling threshold.

\subsection{Software and RAN configuration}
\label{subsec:soft-and-ran-config}

Both gNB stacks are configured with a single Frequency Division Duplex (FDD) cell of 5 MHz bandwidth (25 Physical Resource Blocks, PRBs) and 15 kHz subcarrier spacing (SCS), in FR1 band n1 (UL 1920–1980 MHz, DL 2110–2170 MHz). The sampling rate is 7.68 Msps. %
The cell is configured for terrestrial standalone (SA) operation, without NTN timing parameters or satellite channel emulation. Band n1 is adjacent to the S-band NTN band n256 (UL 1980--2010 MHz, DL 2170--2200 MHz), so the RF chain is representative of an S-band payload. Hence, Sections~\ref{sec:results} and~\ref{sec:ric_results} quantify the onboard cost of the gNB and RIC software, whereas NTN feature support is assessed by code inspection. Section~\ref{sec:lessons} discusses how NTN conditions may affect the results. %
Finally, Table~\ref{tab:sw_versions} lists the software versions.

The experimentation protocol followed consists of the gNB running for a period of time within two and four minutes, traversing the following phases: i) \textit{Boot}, where the gNB initially starts; ii) \textit{Idle}, the cell is on air without UEs; iii) \textit{Attach}, once a UE correctly attaches to the network; and iv) \textit{UL/DL Traffic}, where the UE starts sending or receiving traffic.

\textbf{gNB CPU and memory}. CPU usage is read per thread from \verb|/proc/<pid>/task/<tid>/stat|, differencing user and system ticks once per second. A value of 100\% denotes one Cortex-A53 core fully occupied, so the platform saturates at 400\%. We report both normalizations: per-core values expose single-thread bottlenecks, and platform values show the headroom left for other payload functions to run. We also record voluntary context switches per thread (wake-ups per second), that is, how often a thread blocks and is woken. Memory is the resident and proportional set size (RSS/PSS) from \textit{smaps\_rollup}. Threads are grouped by name into five roles (see Table~\ref{tab:cpu_roles}). Threads that never register a name inherit their creator's name and were identified individually for per-run thread inventories.

\textbf{RIC CPU and memory}. CPU and memory are measured through the control groups of the RIC container, using \textit{cpu.stat} for CPU time and %
\textit{memory.stat} for memory. The values are aggregated over every container of the platform: one for FlexRIC and six for the ORAN-SC RIC. Both gNB and RIC memory figures are therefore RSS-based. The joint footprint described in Section~\ref{sec:ric_results} sums them.

\textbf{Power}. Board power is read every 2s from the \textit{INA226} monitor on the 12V input rail. It covers the whole board, including the DDR memory and the USB-powered radio.

\section{gNB Experimental Evaluation}
\label{sec:results}

\begin{table}
\renewcommand{\arraystretch}{1.15}
\caption{\textbf{Cost per thread role on the ZCU104, averaged over the idle, attach and traffic phases (boot excluded)}}
\label{tab:cpu_roles}
\centering
\small
\begin{tabular}{@{}l r r r | r r r@{}}
\toprule
& \multicolumn{3}{c}{\textbf{OAI}} & \multicolumn{3}{c}{\textbf{srsRAN}} \\
\hline
\cmidrule(lr){2-4}\cmidrule(l){5-7}
\textbf{Thread role} & \textbf{thr.} & \textbf{CPU} & \textbf{wk/s} &
                       \textbf{thr.} & \textbf{CPU} & \textbf{wk/s} \\
\midrule
\hline
RU / low-PHY      &  3 &  59.8\,\% &  7.6\,k &  2 & 111.4\,\% & 28.3\,k \\
Worker pool         &  4 &  54.3\,\% &  3.9\,k &  4 & 109.0\,\% & 38.6\,k \\
SDR / USB &  6 &  42.1\,\% &  7.4\,k &  6 &  27.5\,\% &  5.1\,k \\
L2/L3/control     & 11 &   2.1\,\% &  1.0\,k &  1 &   0.0\,\% &  0.2\,k \\
Main      &  4 &   2.6\,\% &  1.0\,k &  2 &   8.9\,\% &  4.5\,k \\
\midrule
\textbf{Total}      & \textbf{28} & \textbf{160.9\,\%} & \textbf{20.9\,k}
                    & \textbf{15} & \textbf{256.7\,\%} & \textbf{76.6\,k} \\
\bottomrule
\end{tabular}
\end{table}

\subsection{Compute Footprint (CPU and RAM)}

\begin{figure}
  \centering
  \includegraphics[width=0.85\linewidth]{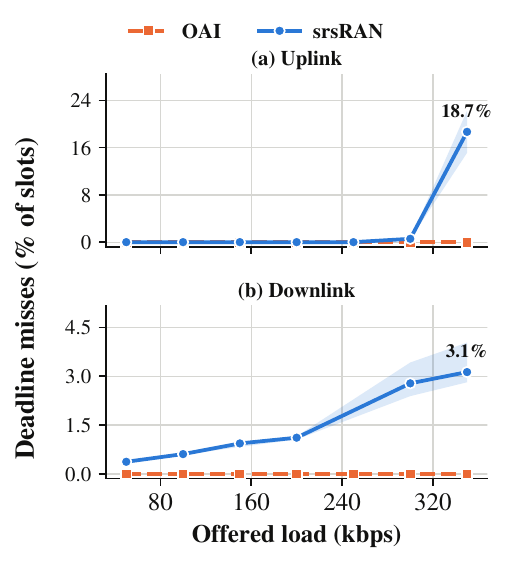}
  \caption{\textbf{Deadline miss rate vs.\ offered load on the ZCU104.}}
  \label{fig:miss_rate}
\end{figure}

Figure~\ref{fig:gnb_cpu} follows both stacks through a complete run and Table~\ref{tab:cpu_roles} resolves the steady state by thread role. The two implementations significantly differ for the same carrier. While srsRAN occupies $\approx260\,\%$ of one core ($\approx66\,\%$ of the platform), 
OAI uses $\approx160\,\%$ (~$\approx42\,\%$ of the platform). 
These values peak at $262.9\,\%$ and $167.7\,\%$ respectively during the UL traffic phase.
This still holds for different data rates under maximum achievable throughput shown in Table~\ref{tab:datarate}. 
Both stacks show high context-switch rates. At $7.68\,\text{Msps}$ and $1\,000$ slots per second, each slot carries $7\,680$ samples: OAI moves them as a single block, whereas srsRAN's lower PHY splits them into $30$ blocks of $256$ samples.  This can be read in Table~\ref{tab:cpu_roles}, where the RU/low-PHY threads wake up (voluntary context switch) $7.6\,\text{k}$ times per second in OAI against $28.3\,\text{k}$ (about $28$ per slot) in srsRAN (in total $20.9\,\text{k}$ against $76.6\,\text{k}$ respectively). 
This means significant thread control overhead for the $1.2\,\text{GHz}$ Cortex-A53 processor. The result is that maximum achievable datarates shown in Table~\ref{tab:datarate} are highly affected by %
processing deadlines and interrupt contention.
Both stacks show maximum datarates in both directions well below what the configured bandwidth and modulation allow. Only the DL datarate in OAI can reach higher values at the expense of disabling HARQ (with HARQ, DL bitrate is limited to 2.75 Mbps). 
To explain these ceilings, Figure~\ref{fig:miss_rate} shows the rate at which the gNB missed slot deadlines as the offered load rises. %
Even for small loads, srsRAN misses deadlines in DL continuously and shows an abrupt increase for UL after 300 kbps. OAI shows no late slots in the same sweep, but its UL ceiling is at 0.55~Mbps. 
However, failure in OAI is different. The stop of UE's PUSCH and PUCCH transmissions coincides with late and underflowed transmit buffers reported by the radio driver.
Since losing those slots also loses the PDCCH carrying the UE's UL grants, we attribute the OAI ceiling to short-term contention between the uplink processing, the transmit path and the USB interrupt handling, %
rather than to sustained CPU load. Section~\ref{sec:decision_guide} discusses how to remove this bottleneck.

Memory reverses the ranking. OAI reaches $667\,\text{MB}$ of resident memory within $15\,\text{s}$ of start and stays flat, because its buffers are pre-allocated for the maximum supported configuration, while srsRAN settles at $376\,\text{MB}$ acquired in steps as the cell comes up and the UE attaches. Both fit the $2\,\text{GB}$ of the board, however OAI's is fixed (predictable before launch) and srsRAN's follows its configured pool and must be re-derived whenever the cell is reconfigured.

\begin{table*}[t]
\centering
\scriptsize
\renewcommand{\arraystretch}{1.25}
\caption{Support of essential NR-NTN features for regenerative LEO operation in OAI (Duranta) and in the srsRAN/OCUDU lineage, by code inspection of the versions indicated. NTN features were not exercised over the air in this work.}
\begin{threeparttable}
\label{tab:ntn_oss_support_essential}
\begin{tabular}{|>{\raggedright\arraybackslash}p{2.2cm}|>{\raggedright\arraybackslash}p{7.4cm}|c|c|c|c|}
\hline
\multirow{2}{*}{\textbf{Category}} & \multirow{2}{*}{\textbf{Essential NTN Feature}} &
\multicolumn{2}{c|}{\textbf{OAI (Duranta) v2.4.0}} & \textbf{srsRAN} & \textbf{OCUDU} \\
\cline{3-6}
 & & \textbf{gNB} & \textbf{nrUE} & \textbf{4bf1543} & \textbf{26.10} \\
\hline

\multirow{2}{*}{\textbf{RF \& PHY}}
 & NTN FR1 bands n254 / n255 / n256 & \cmark & \cmark & \pmark\tnote{a} & \cmark \\ \cline{2-6}
 & Cell-specific $K_\mathrm{offset}$ applied in DL/UL scheduling & \cmark & \cmark & \cmark & \cmark \\
\hline

\multirow{4}{*}{\textbf{Timing \& Sync.}}
 & UE TA pre-compensation (GNSS position + ephemeris) & \na & \cmark & \na & \na \\ \cline{2-6}
 & UE DL Doppler comp. / UL freq. pre-compensation & \na & \pmark & \na & \na \\ \cline{2-6}
 & \texttt{ntn-UlSyncValidityDuration} (Timer T430) & \cmark & \cmark & \cmark & \cmark \\ \cline{2-6}
 & TA reporting (\texttt{ta-Report} in SIB19, TA Report MAC CE) & \xmark & \xmark & \xmark & \cmark \\
\hline

\multirow{7}{*}{\textbf{System Info.}}
 & \texttt{cellBarredNTN} in SIB1 & \cmark & \cmark & \cmark & \cmark \\ \cline{2-6}
 & SIB19 ephemeris: ECEF position/velocity state vector & \cmark & \cmark & \cmark & \cmark \\ \cline{2-6}
 & SIB19 ephemeris: orbital parameters & \xmark & \xmark & \cmark & \cmark \\ \cline{2-6}
 & SIB19 \texttt{epochTime}, \texttt{ta-Common}, \texttt{ta-CommonDrift(Variant)} & \cmark & \pmark\tnote{b} & \cmark & \cmark \\ \cline{2-6}
 & SIB19 \texttt{t-Service}, \texttt{referenceLocation}, \texttt{distanceThresh} & \xmark & \xmark & \xmark & \cmark \\ \cline{2-6}
 & SIB19 NTN neighbour-cell list & \xmark & \xmark & \xmark & \cmark \\ \cline{2-6}
 & Runtime update of SIB19 / \texttt{ntn-Config} during a pass & \pmark & \cmark & \xmark & \cmark \\
\hline

\multirow{4}{*}{\textbf{MAC \& HARQ}}
 & Up to 32 HARQ processes & \cmark & \cmark & \pmark & \cmark \\ \cline{2-6}
 & RTT-aware MAC timers beyond $K_\mathrm{offset}$ & \pmark\tnote{c} & \pmark\tnote{c} & \pmark\tnote{d} & \pmark\tnote{d} \\ \cline{2-6}
 & DL HARQ feedback disabling (\texttt{downlinkHARQ-FeedbackDisabled-r17}) & \pmark\tnote{e} & \cmark & \pmark\tnote{f} & \cmark \\ \cline{2-6}
 & UL HARQ mode B (\texttt{uplinkHARQ-mode-r17}) & \xmark & \xmark & \xmark & \cmark \\
\hline

\multirow{3}{*}{\textbf{Mobility}}
 & Handover with \texttt{ntn-Config} in \texttt{reconfigurationWithSync} & \cmark & \cmark & \xmark & \cmark \\ \cline{2-6}
 & Conditional handover; location-based measurement event D1 & \xmark & \xmark & \xmark & \cmark \\ \cline{2-6}
 & Rel-18 satellite switch with re-synchronization & \xmark & \xmark & \xmark & \cmark \\
\hline

\textbf{Arch. \& Core}
 & Regenerative-specific procedures (ISL over Xn, NG feeder-link switchover) & \xmark & \na & \xmark & \xmark \\
\hline
\end{tabular}
\begin{tablenotes}[flushleft]\scriptsize
\item \cmark: implemented;\; \pmark: partial;\; \xmark: not implemented;\; \na: not applicable to this component.
\item[a] n254 is in the ARFCN table but is classified as FR2, so a 15\,kHz n254 cell is rejected at start-up; fixed in OCUDU.
\item[b] The UE starts the validity timer from the TTI in which SIB19 is received rather than from \texttt{epochTime}.
\item[c] The gNB and nrUE support \texttt{sr-ProhibitTimer-v1700}; DRX is not supported.\quad
\item[d] DRX is supported but modelled without the UE--gNB RTT offset; no \texttt{sr-ProhibitTimer-v1700}
\item[e] \texttt{disable\_harq} sets the IE for all processes, cell-wide, without checking UE capability.
\item[f] \texttt{harq\_mode\_b} reuses DL HARQ processes before feedback arrives, but the IE is not signalled, so the UE keeps sending HARQ-ACK.
\end{tablenotes}
\end{threeparttable}
\end{table*}

\subsection{Storage, Determinism and Boot Time}
The deployable container image of OAI is $589\,\text{MB}$ against
$1\,700\,\text{MB}$ for srsRAN. While less relevant than other metrics, storage competes with mission data, and the volume that must cross the feeder link when the software is updated in
orbit, where a pass lasts minutes. Under \texttt{PREEMPT\_RT}, threads kept their RT priorities, and the misses in Figure~\ref{fig:miss_rate} come from processing capacity and interrupt placement. 
Measured from container start to an on-air cell, OAI is ready in $28\,\text{s}$ and srsRAN in $36\,\text{s}$, matching the boot phases of Figure~\ref{fig:gnb_cpu}. In both stacks, the dominant term is radio bring-up rather than software start-up. However, the two stacks configure the RF front end differently, srsRAN requests a specific master clock rate and pre-allocates USB transfer buffers in both directions, whereas the OAI device arguments set neither. Both figures are short against a LEO visibility window of several minutes, so neither stack constrains a restart within a pass.

\begin{table}[t]
  \centering
  \small
  \caption{\textbf{Achieved out-of-the-box datarate (Mbps) per gNB stack on the ZCU104}}
  \label{tab:datarate}
  \renewcommand{\arraystretch}{1.2}
  \begin{tabular}{cc|cc}
    \toprule
    \multicolumn{2}{c|}{\textbf{OAI}} & \multicolumn{2}{c}{\textbf{srsRAN}} \\
    \midrule
    \hline
    \textbf{DL} & \textbf{UL} & \textbf{DL} & \textbf{UL} \\
    \midrule
    6.9\textsuperscript{a} & 0.55 & 0.47 & 0.43 \\
    \bottomrule
  \end{tabular}
  
  \vspace{4pt}
  {\scriptsize \textsuperscript{a}with HARQ disabled at the gNB (2.75 Mbps with HARQ)}
\end{table}

\subsection{NTN Feature Completeness}
Table~\ref{tab:ntn_oss_support_essential} reviews the main NTN features a regenerative payload deployment needs. Besides the two gNB stacks considered (OAI and srsRAN), we include NTN support for the latest OCUDU~26.10 and the UE from OAI (code inspection only, not benchmarked). Both tested stacks cover the Rel-17 broadcast essentials (NTN FR1 bands, cell-specific $K_{\text{offset}}$, SIB19 ephemeris and timing, T430) and up to $32$ HARQ processes, but differ in several aspects. First, while OAI broadcasts only the state-vector form (Set 1), srsRAN can also broadcast the orbital-element form of the SIB19 ephemeris (i.e., Set 1 and Set 2). Second, OAI signals DL HARQ feedback disabling cell-wide, regardless of UE capability, and lacks UL HARQ mode~B, while srsRAN reuses DL HARQ processes without waiting for feedback, but never signals the UE, so uplink resources are wasted.
Third, only OAI ships an NR-NTN UE for end-to-end prototyping without a commercial NTN terminal. 
Finally, OCUDU closes most srsRAN gaps with per-UE HARQ signalling, %
orbit propagation, conditional handover and the Rel-18 satellite switch, and none of them implements regenerative-specific procedures such as NG feeder-link switchover or ISL transport over Xn.

\section{Near-RT RIC Experimental Evaluation}
\label{sec:ric_results}

Here we provide empirical results for the two near-RT RIC stacks in terms of compute footprint and O-RAN compliance. To do so, we launch the two RICs together with a joint gNB instance in the same ZCU104 that registers and subscribes to them in order to provide an overview of the joint deployment.

\begin{figure}
  \centering
  \includegraphics[width=0.89\linewidth]{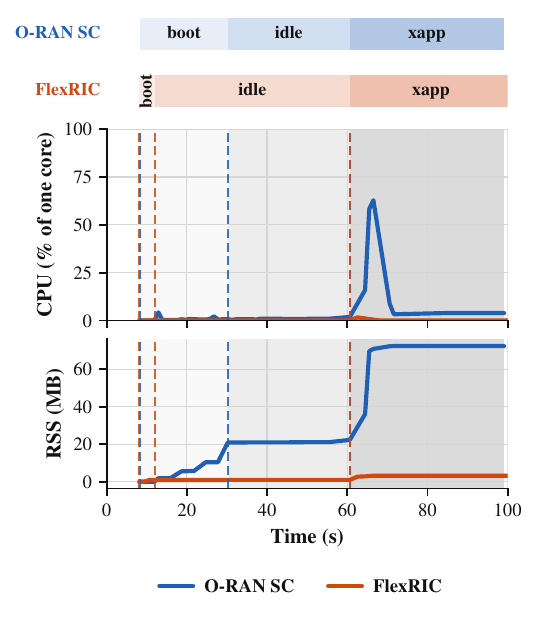}
  \caption{\textbf{Near-RT RIC CPU and RAM usage profile on the ZCU104.}}
  \label{fig:ric_footprint}
\end{figure}

\subsection{Compute Footprint}

\begin{figure}
  \centering
  \includegraphics[width=\linewidth]{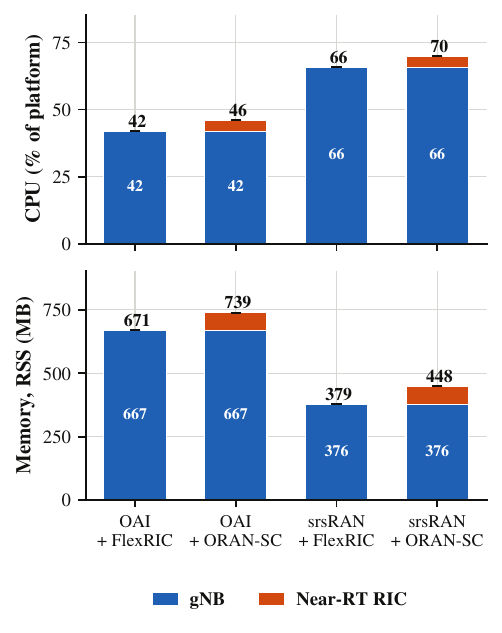}
  \caption{CPU and RAM usage for joint gNB+RIC deployments.}
  \label{fig:joint_footprint}
\end{figure}

In order to show CPU and RAM usage, we deploy sequentially the two RIC stacks along three phases: \emph{boot}, until every service is up; \emph{idle}, with the platform running and no xApp; and \emph{xapp}, with a KPM xApp deployed and subscribed to the E2 agent. This is shown in Figure~\ref{fig:ric_footprint}.  The two platforms have significant differences in terms of processor time and memory. FlexRIC runs as a single process and settles at $3.4\,\text{MB}$ of resident memory and just $0.25\,\%$ of one core.
On the contrary, the ORAN-SC RIC brings up six containers and uses $72\,\text{MB}$ of RAM after the xApp is added, and presents CPU usage transient peaks while onboarding xApps. %
These numbers are shown as joint deployments on the board for the four different combinations in Figure~\ref{fig:joint_footprint}.
On a ZCU104-equivalent payload with 4 Cortex-A53 cores and $2\,\text{GB}$ of memory none of these RIC platforms poses a problem while deployed together with a gNB instance (the gNBs are limited by single-thread performance, not total CPU capacity). Parallel to this, %
there is a difference in the size of their deployable images, $138\,\text{MB}$ for FlexRIC and $2\,381\,\text{MB}$ for ORAN-SC, which is what has to be carried to orbit and re-uploaded when the control logic is updated.

\begin{figure}
  \centering
  \includegraphics[width=\linewidth]{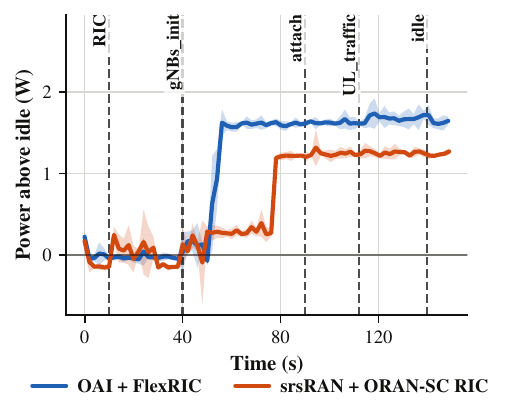}
  \caption{\textbf{Energy profile for two gNB+RIC combinations}}
  \label{fig:power}
\end{figure}

From the energy consumption side, Figure~\ref{fig:power} shows board power for two complete deployments. As an example, we pick OAI+FlexRIC and srsRAN+ORAN-SC.
They go through the same boot, idle, attach and traffic sequence, each averaged over four runs. Both start from an essentially identical idle board ($13.35$ against $13.46\,\text{W}$), and the increments are $+1.66\,\text{W}$ for OAI with FlexRIC and $+1.25\,\text{W}$ for srsRAN with the ORAN-SC RIC. Counterintuitively, the numbers are the opposite of the processor time. The srsRAN-ORAN-SC deployment occupies $70\,\%$ of the platform against $42\,\%$ for the OAI-FlexRIC one, yet draws about 25\% less additional power. 
Occupied processor time may not be a reliable proxy for energy on this experimental setup, which matters because it is the metric most readily available to a payload integrator. %
A plausible hypothesis is that a core stalled on a DRAM access is accounted as busy while the power is drawn by the memory subsystem may be behind this difference in energy consumption. OAI's $667\text{MB}$ needs to compete for the $1\text{MB}$ of shared L2 cache available, while srsRAN holds just $376 \text{MB}$. However, this claim requires further investigation.

\subsection{O-RAN Compliance}

Table 5 summarizes O-RAN compliance for both gNBs and RICs.  The two platforms partially comply on what a Near-RT RIC must do on the gNBs (i.e., both implement E2AP, with some service models such as E2SM-KPM and E2SM-RC). However A1 policy mediation and support for O1 is only provided by ORAN-SC RIC. Also, ORAN-SC RIC isolates xApps in separate containers with their own identities, which increases security at the cost of increasing the resource footprint. Conversely, FlexRIC is remarkably resource efficient and adds custom models for low level metrics and the creation of new E2SMs.  Regarding the E2 interface, all four gNB--RIC combinations completed E2 Setup and delivered KPM indications to a KPM xApp, although OAI with the ORAN-SC RIC required several integration fixes. By default, OAI builds its E2 agent with E2AP v2.03, while the ORAN-SC E2 Termination expects v3.0+, so the SCTP association is established but E2 Setup times out. Building OAI with E2AP v3.0 and E2SM-KPM v3.00 solves this mismatch. 
In addition, the KPM xApp had to match UEs by S-NSSAI, the only matching condition implemented by the OAI's E2 agent.

\begin{table}[t]
\centering
\footnotesize
\setlength{\tabcolsep}{4pt}
\caption{O-RAN compliance of open-source RAN components.}
\begin{threeparttable}
\label{tab:oran_compliance}
\begin{tabular}{@{}lcccc@{}}
\toprule

\textbf{Interface / Feature} & \textbf{srsRAN} & \textbf{OAI} & \textbf{ORAN-SC} & \textbf{FlexRIC} \\
 & \textbf{(OCUDU)} & \textbf{(Duranta)} &  & \\
\midrule
\hline
F1 / E1 (CU--DU split)      & \cmark & \cmark & \na    & \na    \\
Xn                          & \cmark & \xmark & \na    & \na    \\
FAPI (DU-high/DU-low)       & \cmark & \cmark & \na    & \na    \\
Open FH 7.2 (CUS-plane)     & \cmark & \cmark & \na    & \na    \\
Open FH M-plane             & \xmark & \pmark\tnote{a} & \na    & \na    \\
O1                 			& \xmark & \pmark\tnote{b} & \pmark\tnote{c} & \xmark \\
E2AP                        & \cmark & \cmark & \cmark & \cmark \\
E2SM-KPM                    & \cmark & \cmark & \cmark & \cmark \\
E2SM-RC                     & \pmark\tnote{d} & \pmark\tnote{e} & \cmark & \cmark \\
E2SM-CCC         & \pmark\tnote{f}         & \xmark             & \pmark\tnote{g}             & \xmark \\
Custom SMs       & \xmark         & \cmark\tnote{h}         & \xmark             & \cmark\tnote{h} \\
A1                          & \na    & \na    & \cmark & \xmark \\
xApp SDK                    & \na    & \na    & \cmark & \cmark \\

\bottomrule
\end{tabular}
\begin{tablenotes}[flushleft]\scriptsize
\item \item \cmark: implemented;\; \pmark: partial or workaround;\; \xmark: not implemented;\; \na: not applicable to this component.
\item[a] Preliminary validated with specific O-RUs via NETCONF.
\item[b] External O1 adapter (NETCONF/YANG) covering a subset of O1 functions.
\item[c] O1 mediator supports NETCONF (xApp config) and VES (alarms)
\item[d] Control Style~2 only (slice PRB quotas).
\item[e] Report Styles~1/4/5; Control Styles~1/3 (bearer, handover).
\item[f] DU only; O-RRMPolicyRatio (slice PRB) control.
\item[g] Only in srsRAN's ORAN-SC RIC xApp framework, not upstream.
\item[h] MAC/RLC/PDCP/GTP (FlexRIC adds SLICE/TC).
\end{tablenotes}
\end{threeparttable}
\end{table}

\begin{table*}[t]
\renewcommand{\arraystretch}{1.2}
\caption{\textbf{Summary of technical benchmark and trade-off comparison of gNB and Near-RT RIC candidate solutions}}
\label{tab:benchmark_summary}
\centering
\small
\begin{tabularx}{\linewidth}{l | X X X X}
\toprule
\textbf{Metric / KPI} & \textbf{srsRAN/OCUDU (gNB)} & \textbf{OAI/Duranta (gNB)} & \textbf{ORAN-SC (RIC)} & \textbf{FlexRIC (RIC)} \\
\midrule
\hline
\textbf{3GPP NTN Level} & High (gNB, Rel-17/18 LEO/GEO, on-board orbit propagation, SIB19, mobility) & High (nrUE and gNB, Rel-17/18 LEO/GEO, SIB19, timing offsets, doppler compensation) & N/A (Requires xApp) & N/A (Requires xApp) \\
\hline
\textbf{O-RAN Interfaces} & Standard E2 (v3.0+), F1 split, custom minimal O1 & Standard E2 (v2.03 default, v3.0 optional), F1 split, O1 via NETCONF & Full E2 (v3.0+), native A1 Mediator \& O1 & Standard E2 (v2.03, v3.0 optional), custom E2SMs, no A1/O1 \\
\hline
\textbf{Multi-Tenancy} & MOCN, quota-based gNB slicing & MOCN, static/delay-driven MAC slicing & Container/RBAC pod isolation & Dynamic xApp allocations, shared memory \\
\hline
\textbf{CPU (\% of one core)} & {262.9\%} (peak) & {167.7\%} (peak)& {12.3\%} (Spike: 63\%) & {0.25\%} \\
\hline
\textbf{CPU (\% of platform)} & {65.7\%} (peak)& {41.9\%} (peak) & {3.1\%} & {0.06\%} \\
\hline
\textbf{Steady RAM (RSS)} & {376.1 MB} & {667.4 MB} & {72.1 MB} & {3.4 MB} \\
\hline
\textbf{Image Size} & 1700 MB & 589 MB & 2381 MB (6 services) & 138 MB \\
\hline
\textbf{Boot/Recovery Latency} & 36\,s & 28\,s & 18\,s & 2\,s \\
\hline
\textbf{ARM Portability} & Native \texttt{aarch64} build & ISA-specific patches & Substantial container patching & Minor CMake adjustments \\
\hline
\textbf{Licensing} & AGPLv3 / BSD-3-Clause & OAI Public License v1.1 & Apache 2.0 & OAI Public License v1.1 \\
\hline
\textbf{Extensibility} & Moderate (C++ rebuilds) & High (external libraries) & Moderate (Helm/SDK overhead) & High (Lean C/Python SDK) \\
\bottomrule
\end{tabularx}
\end{table*}

\vspace{-1mm}
\section{Discussion and Satellite Payload Selection Guide}
\label{sec:decision_guide}

While the gNB and RIC stacks considered may be each designed for its own purpose, their out-of-the-box performance and functionalities differ significantly. Table~\ref{tab:benchmark_summary} shows an overall summary of the results given in Section~\ref{sec:results} and~\ref{sec:ric_results}. From this table, three main results may drive the selection of the stacks. First,  all combinations fit in the target satellite platform. The gNB dominates the resource footprint (for both CPU and memory). srsRAN uses less memory than OAI, however, at the expense of higher CPU usage. RICs only add a small fraction of that. Second, while in general both gNBs cover the same Rel-17 NTN essentials and implement the main E2AP procedures and E2SMs (i.e., KPM and RC), a closer look at these capabilities and their differences in their resource footprint may clearly tip the decision in favor of one or another component. Third, while FlexRIC offers a highly lean and lightweight deployment, both RICs' footprints are relatively low and the actual decision may be driven by O-RAN compliance or other engineering aspects. In this latter sense, Table~\ref{tab:benchmark_summary} also shows other qualitative metrics, such as ARM portability, licensing or extensibility.

\subsection{Mission Profile Recommendations}

Bearing this in mind, Table~\ref{tab:decision_matrix} provides a decision framework aimed at a class of satellite platforms with a payload processor based on the same MPSoC family as the ZCU104. These recommendations focus on three main mission profiles:  

\begin{itemize}
\item[•] \textbf{Standard High-Capacity LEO Regenerative Payload}: for a single-operator LEO payload, which is the common case for a New Space mission, \textbf{OAI with FlexRIC} may be the most suited baseline. It is the most capable
combination on every axis except memory. It is also the lightest to maintain in orbit (smallest combined image), which bounds the volume of a software update over the feeder link, and the fastest restart for the gNB and the RIC.

\item[•] \textbf{Multi-Tenant Operator-Grade Payload}: a payload shared between operators (through multi-tenancy) may benefit from what only the ORAN-SC RIC provides, i.e., per-xApp container isolation, A1 policy mediation from a non-RT RIC on the ground, and an O1 path into the operator's management system. For this profile, \textbf{OAI with the ORAN-SC RIC} may be the combination to target for LEO (OCUDU for GEO). 
It pairs the more capable gNB with the standards-compliant RIC for a total of about $740\,\text{MB}$ and about $46\,\%$ of the platform. On the other side, the combined container image size approaches
3 GB. However, the ORAN-SC RIC is split into service containers that can be updated separately, so an in-orbit update need not necessarily transfer all of them.

\item[•] \textbf{Severe SWaP Satellite Payload}: in cases where the memory ceiling for the radio function is a few hundred megabytes, and the power budget is very strict, as in small CubeSat, OAI deployments may not fit. Therefore, using the \textbf{srsRAN with FlexRIC} combination, where srsRAN can be configured to use a reduced amount of memory and FlexRIC leaves almost no resource footprint, may allow for a successful deployment in such resource-constrained scenarios.

\end{itemize}

\begin{table*}[t]
\renewcommand{\arraystretch}{1.2}
\caption{\textbf{Satellite payload architecture selection framework}}
\label{tab:decision_matrix}
\centering
\small
\begin{tabularx}{\linewidth}{@{}>{\raggedright\arraybackslash}p{2.7cm} X >{\raggedright\arraybackslash}p{3.2cm} X@{}}
\toprule
\textbf{Mission profile} & \textbf{Primary drivers} & \textbf{Recommended stack} & \textbf{Justification} \\
\midrule
\hline
\textbf{Standard high-capacity LEO regenerative payload} \newline \textit{(single operator)} &
Highly extensible and flexible payload; $58\,\%$ of CPU used available for external functions; 32 HARQ processes cover the RTD,
so feedback disabling is optional; frequent in-orbit software updates. &
\textbf{OAI + FlexRIC} &
Low overall CPU usage that leaves space for external functions (e.g., xApps), ${\approx}67\,\%$ of the memory in the platform is left free. Lightest to maintain in orbit, with a lean RIC, the smallest combined image and the fastest restart. \\
\midrule
\hline
\textbf{Multi-tenant operator-grade payload } \newline \textit{(LEO)}&
Carrier-grade RIC with per-tenant xApp isolation; A1 policy from a ground non-RT RIC; O1 integration with the operator's management plane; $54\,\%$ of CPU used available for external functions &
\textbf{OAI + ORAN-SC RIC} &
Pairs the more capable gNB with the only RIC providing per-xApp container isolation, A1 and O1. Combined image of about 3~GB, but split into service containers that can be updated independently. \\
\midrule
\hline
\textbf{Severe SWaP satellite payload} \newline \textit{(e.g., small CubeSat)} &
Memory ceiling of a few hundred megabytes for the radio function; strict power budget; SIB19 updates
without extra processes; low-rate LEO services. &
\textbf{srsRAN / OCUDU + FlexRIC} &
Smallest memory footprint (${\approx}380\,\text{MB}$), with a configurable buffer pool at the gNB and a lean RIC. Uses ${\approx}66\,\%$ of the board. Link budget may need to be sized for low rates unless optimization are applied. \\
\bottomrule
\end{tabularx}
\end{table*}

\subsection{Lessons Learned and Optimisation Opportunities}\label{sec:lessons}

Every figure in this paper describes the software as it is obtained: upstream
releases, built for \texttt{aarch64} and configured, but not heavily modified. That is
deliberate, because it is the position a mission actually starts from and it
makes the comparison reproducible. It also means the numbers are an upper bound
on cost rather than a limit of the platform, and the experiments point to where
the margin is.

\textit{Core isolation beyond affinity masks.} Both stacks pin their threads with affinity masks inside an otherwise ordinary kernel, so the isolated cores still take timer ticks, RCU callbacks and interrupts. Booting with \texttt{isolcpus}, \texttt{nohz\_full} and \texttt{rcu\_nocbs} over the physical layer cores, and confining device interrupts with \texttt{irqaffinity}, removes that residual work. On our platform, we observed that the GICv2 delivers every USB interrupt to a single core, and both stacks place real-time threads on that same core, so the interrupt that feeds the radio is scheduled behind the threads waiting for it. Correcting the placement is a configuration change with no code impact, and it may be the first step to improve the performance of these measurements.

\textit{Radio attachment and FPGA offload.} 
While the USB transport is not the largest
consumer of processor time (42\,\% of a core in OAI and 28\,\% in srsRAN,
Table~\ref{tab:cpu_roles}), it is where timing determinism is lost. Every slot depends on sample transfers that cross the kernel USB stack and a short stall on the wrong core turns into late transmit buffers and missed uplink receptions, as observed in Section~\ref{sec:results}. Driving an AD9361-class front end directly from the MPSoC
programmable logic (e.g., FMC-attached FMCOMMS3 rather than a USB-attached B210) may remove it entirely, moving timestamping into the FPGA where the slot boundary is a hardware event rather than a scheduling deadline. The same FPGA fabric may even absorb the FFT and LDPC stages if the processor budget still does not close. On a ZCU104-class payload, this hardware is already present and unused.

\textit{Vector kernels for the physical layer.} The two stacks differ in how much of their physical layer is optimized for ARM: srsRAN carries hand-written NEON kernels for its LDPC and modulation paths, while the OAI \texttt{aarch64} build relies more on generic x86 SSE/AVX2 code translated to NEON through the SIMDe library. However at the rates measured here this difference does not show.
Table~\ref{tab:cpu_roles} shows that srsRAN's PHY threads cost more than OAI's despite its native kernels, because at low load the cost is dominated by per-slot processing and thread synchronization (76.6\,k against 20.9\,k wake-ups per second). Still, decoding cost scales with throughput and kernel efficiency will matter once the transport and core-placement limits discussed above are removed and the rate approaches the carrier capacity.

\textit{Right-sizing memory.} OAI's $667\,\text{MB}$ is allocated within $15\,\text{s}$ of start and never grows, because its buffers are dimensioned for
the maximum supported configuration rather than the $25$ PRBs actually in use. Sizing them to the deployed cell is a build-time change that would bring OAI within reach of the memory ceiling that currently excludes it from a small CubeSat profile, and would reduce pressure on a memory subsystem that may cost more energy than the cores themselves. srsRAN already exposes its buffer pool as a configuration parameter, which is why its footprint may well track the cell it is serving.

\textit{Slimming the RIC.} The ORAN-SC deployment runs six services because that is what its composer file starts, not because a payload needs all of them. Dropping %
unnecessary components may significantly further reduce not only the CPU and memory usage but also the container image size, which is a figure that matters for an in-orbit software update.

\textit{Moving to real NTN conditions.} Enabling NTN operation would change the measured results in several ways. First, the cell-specific $K_{\text{offset}}$ shifts the uplink timing relationship. srsRAN needs a longer \texttt{max\_proc\_delay} than the random-access response window allows (a value higher than 2 exceeds the maximum of 10 slots), because both are counted in slots against the same DL reference. An NTN configuration adds $K_{\text{offset}}$ to this timing, which is precisely the slack the srsRAN worker pool lacks. Second, disabling feedback removes retransmissions, and may raise the  DL rate (e.g., see OAI DL increasing from 2.75 to 6.9\,Mbps in Table~\ref{tab:datarate}), however at the cost of relying on RLC for reliability. Third, the lower SNR of a satellite link forces lower MCS and more LDPC decoding iterations, which may increase the uplink physical-layer compute. Finally, with the Near-RT RIC on board, the E2 loop does not traverse the feeder link with only A1 and O1 exposed to the ground-segment delay. Validating these effects with NTN timing enabled and an emulated satellite channel, using the OAI nrUE, is left for future work.

\textit{Generalization to other platforms and configurations.} The results were obtained on a single platform with a $5\,\text{MHz}$ carrier, and not all of them generalize equally. Relative results, i.e., the gNB dominating the joint footprint, the Near-RT RIC adding only a few percent of CPU, the memory ranking of the gNBs, image sizes and radio-dominated boot times, stem from the software architecture and should hold on other ARM payload processors. However, absolute CPU usage and datarates for the $1.2\,\text{GHz}$ in-order Cortex-A53 are conservative baselines, and other out-of-order cores should offer more single-thread margin, which is the resource the gNBs seem to lack. Regarding bandwidth, the sampling rate and the FFT and LDPC workload grow roughly linearly with the number of PRBs ($15.36$ and $30.72\,\text{Msps}$ for $10$ and $20\,\text{MHz}$ respectively), so, given the load measured at $5\,\text{MHz}$, wider carriers on this class of processor would require the FPGA offload and core isolation discussed above. %

\section{Conclusions}
\label{sec:conclusion}

This paper benchmarked two open-source gNB stack families, OAI/Duranta and srsRAN/OCUDU, and two Near-RT RICs, the ORAN-SC RIC and FlexRIC, deployed out of the box on a ZCU104 whose MPSoC is commonly used in already flying small-satellite payloads. Onboard O-RAN on this class of hardware is feasible
today, and every gNB--RIC combination fits within the platform. The gNB dominates the budget, at $42$ to $66\,\%$ of the platform and $376$ to
$667\,\text{MB}$, whereas a Near-RT RIC adds at most a few percent of CPU and $72\,\text{MB}$, so closed-loop control can be hosted on board at little cost. %
These results lead to three recommendations: OAI with FlexRIC for high-capacity single-operator payloads, OAI with the ORAN-SC RIC for multi-tenant payloads, %
and srsRAN/OCUDU with FlexRIC where memory and power are the binding constraints. The results presented are bounded by a single platform with a USB-attached COTS front
end, a terrestrial configuration without NTN timing or channel emulation. We foresee that applying the several optimization strategies provided, bitrates may significantly increase to offer carrier-grade NTN services. %

\acknowledgements

This work was partially supported by the European Space Agency (ESA) under the ARTES Space Systems for Safety and Security (4S) programme, contract No. 4000150317/25/NL/GM/fm (activity 3F.070), project ``Secure Real-Time OS on SW-Defined Radio for Satellite-based 5G Radio Access Network" (SULFUR), with the support of the Spanish Space Agency (Agencia Espacial Española, AEE). The view expressed herein can in no way be taken to reflect the official opinion of the European Space Agency. It has also been co-funded by Departament de Recerca i Universitats under the frame of the Resilicat project.

\IfFileExists{IEEEtran.bst}{\bibliographystyle{IEEEtran}}{\bibliographystyle{ieeetr}}
\bibliography{references}

\thebiography
\begin{biographywithpic}
{Esteban Municio}{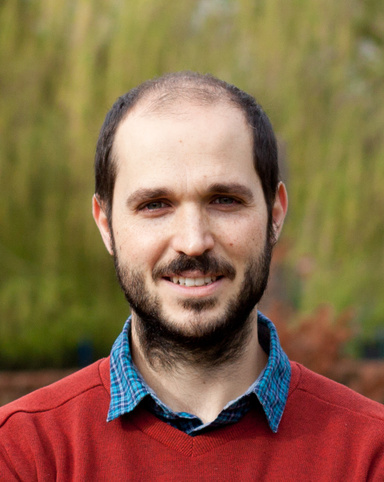} is a Telecommunication Engineer and Senior Researcher
specialized in wireless networks currently working in the AI-driven Systems group of the i2CAT Foundation. He obtained his Ph.D. degree from the
University of Antwerp in 2020 within imec’s research group IDLab.
He then continued in imec as postdoctoral researcher for two more years.
Since January 2022, he has been with i2CAT, where currently he is Senior
Researcher at the AI-driven Systems group. He is currently Adjunct Professor
at Universidad Carlos III de Madrid (UC3M) in the Master in Connected
Industry 4.0. He has been working in a number of EU projects within FP7,
H2020, Horizon Europe and ESA-ARTES. His research interests are in the field of AI-driven programmable networks, Open RAN, TSN, ultra-reliable Industrial IoT and Non-Terrestrial Networks. He is also interested in
testbed deployments, community networks and connectivity provision in rural
environments.
\end{biographywithpic}

\begin{biographywithpic}
{Oscar Baselga}{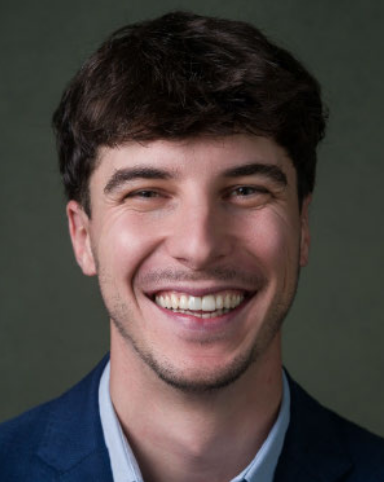}
was born in Zaragoza, Spain.
He received the B.S. degree in computer engineering from the Universidad de Zaragoza (UZ),
Spain, in 2021, and the M.S. degree in the Internet
of Things from the Universidad Complutense de
Madrid (UCM), Spain, in 2022. He is currently
pursuing the Ph.D. degree in network engineering with the Universitat Politécnica de Catalunya
(UPC), Spain. He is also a First Stage Researcher
and a Ph.D. Candidate with the Space Communications Research Group, i2CAT Research Centre,
Barcelona, Spain. His professional experience centers on the development
of 5G NTN technologies for regenerative satellites.
His research interests include functional split architectures, the optimization
of on-board gNB implementations, space multi-tenant RAN, and space-based O-RAN solutions.
\end{biographywithpic}

\begin{biographywithpic}
{Oscar Lasierra}{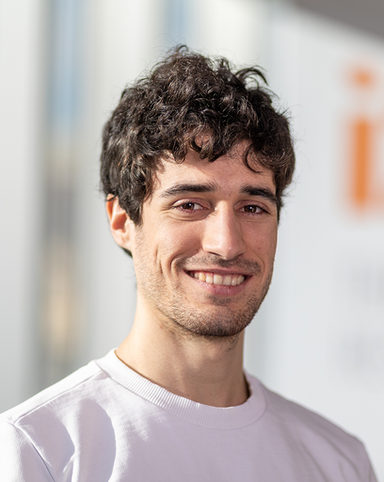}
received the B.Eng. degree in telecommunications
from the Polytechnic University of Catalonia, Castelldefels. In 2022, he
joined i2CAT as a Researcher with the AI-Driven Systems Group.
His research interests encompass a wide range of topics, including software-
defined radio, cellular networks, and security, and privacy.
\end{biographywithpic}

\begin{biographywithpic}
{Gines Garcia-Aviles}{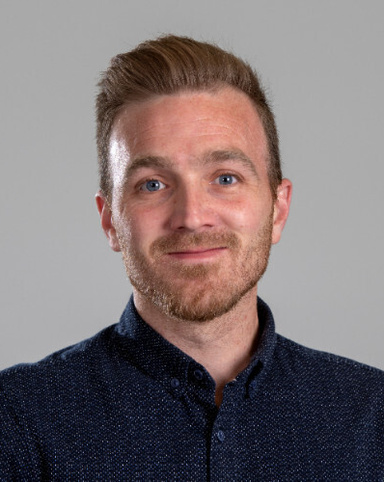}
received his M.Sc. (2018) and PhD (2021, cum laude with International Mention) in Telematics Engineering from the University Carlos III of Madrid, Spain, where his dissertation received the Extraordinary Doctorate Award. During his doctoral studies, he was part of the wireless networking group at IMDEA Networks Institute and the Telematics Department at UC3M. In 2021 he joined the i2CAT Foundation as a post-doctoral researcher. From 2024 to 2026, he was a post-doctoral researcher at the University of Murcia, Spain, lecturing in mobile computing and cybersecurity. He has contributed to numerous EU-funded 5G/6G projects. His research interests include RAN virtualization and O-RAN, network slicing and orchestration, 5G/6G security, and AI-driven network management.
\end{biographywithpic}

\begin{biographywithpic}
{Xavier Costa-Pérez}{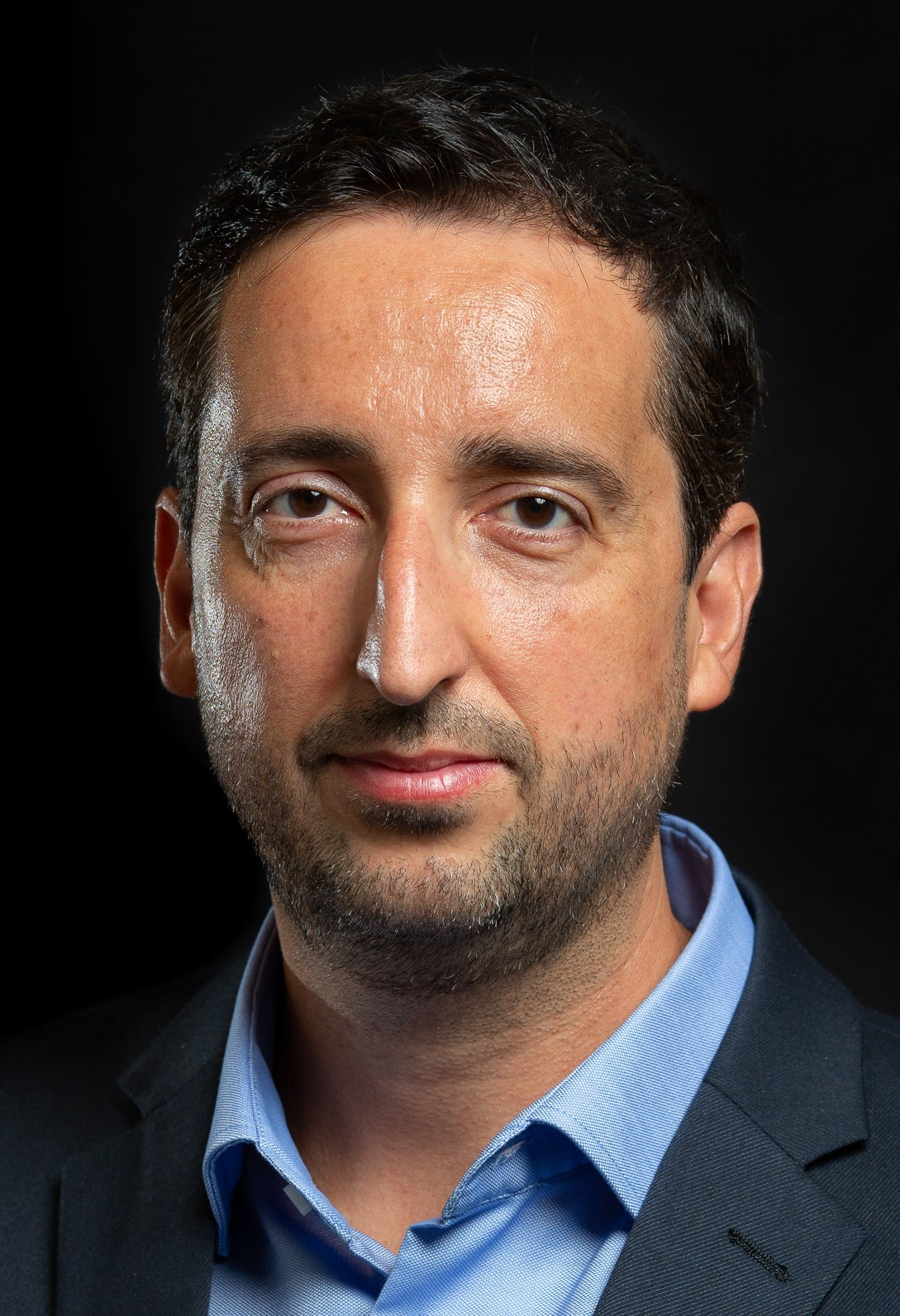}
is an ICREA Research Professor and Scientific Director at the i2cat Research Center, concurrently leading Space and Terrestrial Systems (ASTRO) R\&D at NEC Laboratories Europe. He has been ranked among the Top 2\% of World Scientists by Stanford for over 5 years and is featured on the Research.com Best World Scientists list. His innovations have been integrated into commercial mobile phones, base stations, and network management systems, and have spurred the creation of multiple start-ups. He has served on the organizing committees of AAAI, ACM MobiCom and IEEE INFOCOM, and as an Editor of IEEE TMC, IEEE TCOM and Elsevier COMCOM. Dr. Costa-Pérez holds M.Sc. and Ph.D. degrees in Telecommunications from the Polytechnic University of Catalonia (UPC), receiving a national award for his doctoral thesis. He is the inventor of over 100 granted patents, some standard essential (SEP).
\end{biographywithpic}

\end{document}